\documentclass[aps,twocolumn]{revtex4}
\usepackage{amsmath} \usepackage{amsfonts} \usepackage{amssymb}

\begin{document}

\title{On the origin of the difference between time and space}

\author{C. Wetterich}

\address{
Institut f{\"u}r Theoretische Physik,
Philosophenweg 16, 69120 Heidelberg, Germany}


\begin{abstract}
We suggest that the difference between time and space is due to spontaneous symmetry
breaking. In a theory with spinors the signature of the metric is related to the
signature of the Lorentz-group. We discuss a higher symmetry that contains
pseudo-orthogonal groups with arbitrary signature as subgroups. The fundamental
asymmetry between time and space arises then as a property of the ground state rather
than being put into the formulation of the theory a priori. We show how the complex
structure of quantum field theory as well as gravitational field equations arise from
spinor gravity - a fundamental spinor theory without a metric.

{~}

\end{abstract}
\maketitle


\vspace{1.0cm}

In special and general relativity time and space are treated in a
unified framework. Nevertheless, a basic asymmetry between these two
concepts persists, related to the signature of the metric. It is at
the root of much of the complexity of physics and the universe. The
quantum field equations with a Euclidean signature often admit as
solutions only a single ground state or a few (sometimes degenerate)
states. In contrast, the Minkowski signature allows for many complex
solutions with causal time evolution. Within the presently existing
attempts to find a unified theory of all interactions, based on
quantum field or superstring theories, this time-space asymmetry is
assumed a priori in the form of a given signature.

In this letter we pursue the perhaps radical idea that the difference
between time and space arises as a consequence of the ``dynamics'' of
the theory rather than being put in by hand. More precisely, we will
discuss a model where the ``classical'' or ``microscopic'' action does
not make any difference between time and space. The time-space
asymmetry is generated only as a property of the ground state and can
be associated to spontaneous symmetry breaking.

In a model with only bosonic fields the idea of associating the
Minkowski signature to a ground state property may seem
straightforward. One could discuss a quantum theory for a symmetric
second rank tensor (metric) field which can take arbitrary positive or
negative values for its elements. The expectation value of the metric
defines then its signature and the ground state could indeed
correspond to a Minkowski signature for an appropriate model. The
situation  changes drastically, however, in the presence of fermionic
fields. The coupling of fermions to the gravitational degrees of
freedom involves the vielbein $e^m_\mu$. Both the spinors and the
vielbein transform nontrivially under the ``Lorentz group''. In turn,
the connection between the vielbein and the metric, $g_{\mu \nu}=e^m_\mu
e^n_\nu\eta_{mn}$, involves the invariant tensor $\eta_{mn}$ of this
Lorentz group. The signature of the metric is then uniquely related to
the signature of $\eta_{mn}$ and therefore fixed once the Lorentz
group is specified.

In consequence, a fermionic model for the spontaneous generation of
the time-space asymmetry should not have a  ``bias'' for a particular
signature of the Lorentz group. We will discuss actions that are
invariant under the complex orthogonal group $SO(d,{\mathbb C})$. This
non-compact group admits as subgroups all pseudo-orthogonal groups
$SO(s,d-s)$ with arbitrary signature $s$.  However, the ground state
may spontaneously break this symmetry, being  invariant only with
respect to pseudo-orthogonal transformations with a given $s$. Within
our particular 16-dimensional  model we will even give a reason why
the ground state may favor the ``physical signature'' $s=1$ as
compared to the Euclidean signature $s=0$.

Our approach is based on spinor gravity - a theory involving only
fundamental fermions and no fundamental metric \cite{Aka}-\cite{Wet}.
Nevertheless, the action is invariant under general coordinate
tranformations (diffeomorphisms). The vielbein and the metric appear
as composite objects involving two or four fermions.
As a specific model we consider a $16$-dimensional theory where the
fermions are represented by $256$ Grassmann variables
$\psi_\alpha(x^\mu),~\alpha=1\dots 256,~\mu=0\dots 15,$
$\big\{\psi_\alpha(x),\psi_\beta(y)\big\}=0$. Below, we will construct an
action $S[\psi]$ as an element of a real Grassmann algebra, i.e. as a ``sum''
(or integral) of polynomials in $\psi$ with real coefficients. In a
first step we will require that the
action is invariant under the group $SO(128,{\mathbb C})$ generated by
the infinitesimal transformations
\begin{equation}\label{1}
\delta\psi=\left(\begin{array}{rr}\rho,&-\tau\\\tau,&\rho\end{array}\right)\psi
\end{equation}
with $128\times 128$ blocks of real antisymmetric matrices $\rho=-\rho^T,\tau=-\tau^T$.
This is a huge non-compact group with $128\cdot 127$ real generators.

The group
$SO(128,{\mathbb C})$ has as subgroups all pseudoorthogonal groups $SO(s,16-s)$
for arbitrary signature $s$. Indeed, we may first embed $SO(16,{\mathbb C})$ into
$SO(128,{\mathbb C})$ by restricting $(m,n\!=\!0\dots 15,\epsilon_{mn}\!=\!-\epsilon_{nm},
\bar{\epsilon}_{mn}\!=\!-\bar{\epsilon}_{nm})$
\begin{equation}\label{2}
\rho=-\frac{1}{2}\epsilon_{mn}\hat{\Sigma}^{mn}~,~
\tau=\frac{1}{2}\bar{\epsilon}_{mn}\hat{\Sigma}^{mn}.
\end{equation}
Here $\hat{\Sigma}^{mn}=-\hat{\Sigma}^{nm}$ are the generators of $SO(16)$ in
the $128$-component Majorana-Weyl representation \cite{Wet2}.
 They are represented by real antisymmetric
$128\times 128$-matrices. (For $SO(d)$ the real $2^{\frac{d}{2}-1}$-dimensional spinor
representation exists only if $d=8~mod~8$.) The $240$ real generators of $SO(16, {\mathbb C})$
can be taken as
\begin{eqnarray}\label{3}
\Sigma^{mn}_E&=&\hat{\Sigma}^{mn}{\, 1}~,~B^{mn}=-\hat{\Sigma}^{mn}I,\nonumber\\
I&=&\left(\begin{array}{rr}0&-1\\1&0\end{array}\right),~I^2=-1.
\end{eqnarray}
They obey the commutation relations
\begin{eqnarray}\label{4}
\left[\Sigma^{mn}_E,\Sigma^{pq}_E\right]&=&f^{mnpqst}\Sigma^{st}_E,\nonumber\\
\left[B^{mn},B^{pq}\right]&=& -f^{mnpqst}\Sigma^{st}_E,\nonumber\\
\left[\Sigma^{mn}_E,B^{pq}\right]&=&f^{mnpqst}B^{st}
\end{eqnarray}
with $f^{mnpqst}$ the usual structure constants of $SO(16)$ (with $(mn)$ etc.
considered as double-index). The pseudo-orthogonal group $SO(s,d-s)$ with
signature $s$ obtains by taking  $\epsilon^{(s)}_{mn}=\epsilon_{mn}$
for $m\geq s,n\geq s$,
$\epsilon^{(s)}_{mn}=-{\epsilon}_{mn}$ for $ m<s,n<s$ and
$\epsilon^{(s)}_{mn}=\bar{\epsilon}_{mn}$ otherwise, i.e. if one index
is smaller and the other larger or equal $s$.

At this point everything is formulated within a real Grassmann algebra and no
particular signature is singled out - the different $s$ just denote different
subgroups of the symmetry group of the action, $SO(128, {\mathbb C})$.
 Within our spinor
theory the ground state may be characterized by non-vanishing expectation values of
(bosonic)
spinor bilinears or, more generally, by composites with even powers of Grassmann variables.
 Let us consider the bilinears
\begin{equation}\label{5}
\tilde{E}^0_\mu=\psi_\alpha\partial_\mu\psi_\alpha~,~
\tilde{E}^k_\mu=\psi_\alpha(\hat{a}^kI)_{\alpha\beta}
\partial_\mu\psi_\beta
\end{equation}
where $\hat{a}^k$ denote $15$ real antisymmetric $128\times 128$-matrices obeying the
anticommutation relation
\begin{equation}\label{6}
\left\{\hat{a}^k,\hat{a}^l\right\}=-2\delta^{kl}~,~
k,l=1\dots 15.
\end{equation}
Such matrices $\hat{a}^k$ exist since $\hat{\gamma}^k=i\hat{a}^k$ spans the
Clifford algebra which admits for $d=15$ a $128$-component spinor representation
with purely imaginary and antisymmetric $\hat{\gamma}^k$ \cite{Wet2}.
 We can use $\hat{a}^k$ for the
construction of the $SO(16)$ generators
\begin{equation}\label{7}
\hat{\Sigma}^{kl}=\frac{1}{4}[\hat{a}^k,\hat{a}^l]~,
~\hat{\Sigma}^{0k}=-\frac{1}{2}\hat{a}^k
\end{equation}
with $\hat{\Sigma}^{01}\hat{\Sigma}^{23}\dots \hat{\Sigma}^{14,15}=1/256$.
In particular, it is obvious that $\tilde{E}^k_\mu$ transforms under global $SO(15)$ as
a $15$-dimensional vector whereas $\tilde{E}^0_\mu$ is a singlet.

Actually, the bilinear $\tilde{E}^m_\mu$ transforms under global $SO(1,15)$ as a
$16$-dimensional Lorentz-vector. This can be seen by writing
\begin{eqnarray}\label{8}
\tilde{E}^m_\mu&=&\bar{\psi}\beta^m\partial_\mu\psi~,~\bar{\psi}=\psi^T\beta^0~,\nonumber\\
\beta^0&=&\left(\begin{array}{cc}0&1\\1&0\end{array}\right),
\beta^k=\beta_0\hat{a}^kI
\end{eqnarray}
where
\begin{eqnarray}\label{9}
\{\beta^m,\beta^n\}&=&-2\eta^{mn}_M~,~[\beta^m,\beta^n]=\frac{1}{4}\Sigma^{mn}_M\nonumber\\
\eta^{mn}_M&=&diag(-1,1\dots 1),~ \nonumber\\
\Sigma^{0k}_M&=&\frac{1}{2}\hat{a}^kI=B^{0k}~,~\Sigma^{kl}_M=\Sigma^{kl}_E.
\end{eqnarray}
We observe that for an $SO(1,15)$ Lorentz-transformation
\begin{equation}\label{10}
\delta_{\cal L}\psi=-\frac{1}{2}\epsilon^M_{mn}\Sigma^{mn}_M\psi~,~
\delta_{\cal L}\bar{\psi}=\frac{1}{2}\epsilon^M_{mn}\bar{\psi}
\Sigma^{mn}_M
\end{equation}
one obtains indeed
\begin{equation}\label{11}
\delta_{\cal L} \tilde{E}^m_\mu=-\tilde{E}^n_\mu(\epsilon^M)_n\ ^m~,~
(\epsilon^M)_n\ ^m=\epsilon^M_{np}\eta^{pm}_M.
\end{equation}
We emphasize that our model does not allow the construction of the vector representation
of $SO(16)$ from spinor bilinears. Only $15$ matrices $\hat{a}^k$ obeying eq. (\ref{6})
exist. On a group theoretical level the vector representation is contained in the
product of two inequivalent irreducible spinor representations of
$SO(s,16-s)$. From eq. (\ref{1}) we see that for $SO(128)$ (as
represented by $\rho$) 
and all its
subgroups including $SO(16)$ the fermion field $\psi$ is reducible into two
{\it identical} real $128$-dimensional representations. In contrast, for $s=1$ the
$256$-component Majorana spinor is equivalent to a $128$-dimensional complex Weyl-spinor
which is not equivalent to its complex conjugate. The product of two
spinors therefore contains a vector for $SO(1,15)$ but not for $SO(16)$.

This simple algebraic property could be the root\footnote{Ground states with $SO(16)$
symmetry are possible if bilinears with a different tensor struture condense. Also
$SO(15)$ or subgroups of it like $SO(4)$ can be left invariant by suitable expectation
values of $\tilde{E}^m_\mu$.} of the observed asymmetry between time and space\,!
Indeed, consider an expectation value of the form
\begin{equation}\label{12}
E^m_\mu(x)=\langle\tilde{E}^m_\mu(x)\rangle=\delta^m_\mu.
\end{equation}
It is left invariant if the global Lorentz transformation (\ref{10}), (\ref{11}) is
accompanied by a suitable general coordinate transformation
\begin{equation}\label{13}
\delta_\xi E^m_\mu(x)=-\xi^\nu(x)\partial_\nu E^m_\mu(x)-\partial_\mu
\xi^\nu(x) E^m_\nu(x).
\end{equation}
For a global Lorentz transformation of the coordinates the ground state
(\ref{12}) remains indeed invariant
\begin{equation}\label{14}
\xi^\nu=(\epsilon^M)^\nu\ _\mu x^\mu~,~(\delta_\xi+\delta_{\cal L})E^m_\mu=0.
\end{equation}
The transformation property (\ref{11}) carries over to the global vielbein
$E^m_\mu(x)=\langle\tilde{E}^m_\mu(x)\rangle$ and we can construct a metric
which is invariant under $\delta_{\cal L}$ by using the $SO(1,15)$ invariant tensor
$\eta_{M,mn}$
\begin{equation}\label{15}
g_{\mu\nu}(x)=E^m_\mu(x)E^n_\nu(x)\eta_{Mmn}.
\end{equation}
A ground state $g_{\mu\nu}=\eta_{M\mu\nu}$ is then invariant under the coordinate
transformation (\ref{14}).

Let us next construct an action $S$ for spinor gravity according to the following
principles: i)$S$ is an element of a real Grassmann algebra,
(ii) $S$ is invariant under general coordinate transformations (diffeomorphisms)
$\delta_\xi\psi=-\xi^\nu\partial_\nu\psi$,
(iii) $S$ is invariant under global $SO(128,{\mathbb C})$ transformations
(\ref{1}), (iv) $S$ is local. Under these conditions we will find that $S$
can only be a sum of six invariants which involve either $144$ or $274$ powers of
$\psi$. We also find that $S$ is actually invariant under {\em local}
$SO(128, {\mathbb C})$ transformations if two of the six couplings vanish.

For this construction it is useful to exploit the complex structure which is compatible with
the transformation (\ref{1}). Complex conjugation can be associated to an involution
$\psi_{128+\hat{\alpha}}\rightarrow-\psi_{128+\hat{\alpha}},\hat{\alpha}=1\dots 128$, i.e.
\begin{equation}\label{16}
\psi\rightarrow K\psi~,~K=\left(\begin{array}{rr}1&0\\0&-1\end{array}\right)~,~
K^2=1.
\end{equation}
We can map $K$-odd quantities onto purely imaginary variables and introduce an $128$-
component {\em complex} Grassmann variable
\begin{equation}\label{17}
\varphi_{\hat{\alpha}}=\psi_{\hat{\alpha}}+i\psi_{128+\hat{\alpha}}~,~
\varphi_{\hat{\alpha}}^*=\psi_{\hat{\alpha}}-i\psi_{128+\hat{\alpha}}.
\end{equation}
The transformation (\ref{1}) can now be written as a complex matrix multiplication
\begin{equation}\label{18}
\delta\varphi_{\hat{\alpha}}=\sigma_{\hat{\alpha}\hat{\beta}}\varphi_{\hat{\beta}}~,~
\sigma=\rho+i\tau
\end{equation}
which is compatible with the complex structure, $\delta\varphi^*=\sigma^*\varphi^*$.
The $\sigma$ are arbitrary complex antisymmetric $128\times 128$ matrices, explaining
the name $SO(128,{\mathbb C})$. Invariants under global $SO(128,{\mathbb C})$
transformations involve an appropriate number of $\varphi_{\hat{\alpha}}$ contracted
with the two invariant tensors $\delta^{\hat{\alpha}\hat{\beta}}$ or
$\epsilon^{\hat{\alpha}1\dots\hat{\alpha}_{128}}$ - in contrast, contractions
of mixed terms involving $\varphi$ and $\varphi^*$ are not invariant. Diffeomorphism
symmetry is realized if precisely $16$ derivatives of $\varphi$ are contracted with
$\epsilon^{\mu_1\dots \mu_{16}}$ and we find the general form of the invariant action
\begin{eqnarray}\label{19}
S&=&\alpha\int d^dxW[\varphi]R(\varphi,\varphi^*)+c.c.,\nonumber\\
W[\varphi]&=&\frac{1}{16!}\epsilon^{\mu_1\dots\mu_{16}}
\partial_{\mu_1}\varphi_{\hat{\alpha}_1}\dots\partial_{\mu_{16}}
\varphi_{\hat{\alpha}_{16}}L^{\hat{\alpha}_1\dots\hat{\alpha}_{16}}.
\end{eqnarray}
Here $L$ is the totally symmetric invariant tensor of rank $16$
\begin{equation}\label{20}
L^{\hat{\alpha}_1\dots\hat{\alpha}_{16}}=
sym\left\{\delta^{\hat{\alpha}_1\hat{\alpha}_2}
\delta^{\hat{\alpha}_3\hat{\alpha}_4}\dots
\delta^{\hat{\alpha}_{15}\hat{\alpha}_{16}} \right\}
\end{equation}
and $sym$ denotes total symmetrization in all indices. The remaining part
$R(\varphi,\varphi^*)$ is a local polynomial not involving derivatives.
Due to the anticommuting properties the contractions can only involve the
$\epsilon$-tensor
\begin{eqnarray}\label{21}
R(\varphi,\varphi^*)=T(\varphi)+\tau T(\varphi^*)+\kappa T(\varphi)T(\varphi^*),\nonumber\\
T(\varphi)=\frac{1}{128!}\epsilon^{\hat{\beta}_1\dots\hat{\beta}_{128}}
\varphi_{\hat{\beta}_1}\dots\varphi_{\hat{\beta}_{128}}.
\end{eqnarray}
(For $R=1$ the action (\ref{19}) becomes a total derivative.) We note
($\big(\alpha
W[\varphi]T(\varphi)\big)^*=\alpha^*W[\varphi^*]T(\varphi^*)$ etc.)
 that by use of eq. (\ref{17}) the action (\ref{19}) can be
explicitely written as an element of a real Grassmann algebra for the variables
$\psi_\alpha$.

The action is characterized by three dimensionless complex (or six real) coupling constants
$\alpha,\tau$ and $\kappa$. For $\tau=0$ it is invariant under {\em local}
$SO(128,{\mathbb C})$ transformations. Due to the identity
$\varphi_{\hat{\alpha}}(x)\varphi_{\hat{\alpha}}(x)=0$ (no summation over
$\hat{\alpha}$ here, also note
$\hat{\varphi}_\alpha\hat{\varphi}_\alpha^*=2i\psi_{128+\hat{\alpha}}
\psi_{\hat{\alpha}}\neq 0)$ at most $128$ powers of $\varphi$ can occur at a given location
$x$ such that the inhomogeneous contribution from $\delta_{\cal
  L}(\partial_\mu\varphi)$ 
does not
contribute to $\delta_{\cal L}S$.
(For details see \cite{Wet}.)
 Local Lorentz symmetry, if free of anomalies, has important
consequences for the spectrum of gravitational excitations. Indeed, a
model with only global Lorentz symmetry leads to additional massless
gravitational degrees of freedom \cite{Heb,Wet}.
We concentrate here on
the case of local $SO(128,{\mathbb C})$ symmetry
with $\tau=0$. Imposing an additional discrete symmetry can further reduce
the number of allowed couplings. In particular, the transformations
$\varphi_1\leftrightarrow\varphi_2$ or $\varphi_1\rightarrow-\varphi_1$ are not part of
$SO(128,{\mathbb C})$ and map
$ W\leftrightarrow  W,T\leftrightarrow-T$.
The reflection of one coordinate $x^0\rightarrow -x^0$ results in
$ W\leftrightarrow- W,T\leftrightarrow T$. Requiring invariance under the combined
transformation restricts $\kappa=0$. Finally, a reflection symmetry
$\varphi\leftrightarrow\varphi^*$ (\ref{16}) maps
$ W\leftrightarrow W^*,T\leftrightarrow T^*$ and would
leave only one real coupling $\alpha$. This can be scaled to an
arbitrary value (e.g. $\alpha=1$) by an appropriate scaling of $\psi$.

The existence of only a small number of allowed dimensionless
 couplings strongly suggests that the
theory may be renormalizable! After a proper regularization of the model
(i.e. specification of a well defined functional measure) the couplings may evolve in
dependence on some appropriate ``renormalization scale'' $\mu$. The existence of a
(partial) infrared fixed point would further enhance the predictive
power of the model
even for the case of several couplings.
If the partial fixed point has precisely one marginal direction the corresponding
dimensionless coupling would behave similar to the gauge coupling in QCD and could generate
a characteristic mass scale by dimensional transmutation. All dimensionless quantities
(like ratios of mass scales) would become, in principle, predictable,
similar to the case of a single real coupling $\alpha$. We also note that more
invariants (with less powers of $\psi$) become possible if the symmetry of the action
is only a subgroup of $SO(128, {\mathbb C})$, like $SO(16,{\mathbb C})$, which would be
sufficient for our purpose.

The action $S(19)$ can be viewed as an element of a complex Grassmann
algebra. Complex conjugation is defined by $\varphi\to\varphi^*$,
accompanied by a complex conjugation of all coefficients. (This is the
meaning of c.c. in eq. (\ref{19})  such that $S^*=S$.) Hermitean
conjugation involves an additional transposition of all Grassmann
variables. We can extend this to a more general form of the action and
note $S^\dagger =-S^*$ if $S$ is a sum of elements with $n=2~mod~4$
powers of $\varphi$ or $\varphi^*$, whereas $S^\dagger=S^*$ for
$n=4~mod~4$.
The bilinear $(k\not= 0)$
\begin{eqnarray}\label{21AA}
\tilde E^0_\mu&=&\frac{1}{2}(\varphi^\dagger
\partial_\mu\varphi+\varphi^T\partial_\mu\varphi^*),\nonumber\\
\tilde E^k_\mu&=&\frac{i}{2}(\varphi^\dagger\hat
a^k\partial_\mu\varphi-\varphi^T\hat a^k\partial_\mu\varphi^*)
\end{eqnarray}
is antihermitean and real. The Dirac matrices $\gamma^m=i\beta^m$ are
purely imaginary and obey $\{\gamma^m,\gamma^n\}=2\eta^{mn}_M$.

The generating functionals for the vielbein and its (connected)
correlation functions are defined as 

\begin{equation}\label{22}
Z[J]=\int{\cal D}\psi\exp\left\{-S_E+\int dx\,J^\mu_m(x)\tilde
  E^m_\mu(x)\right\}
\end{equation}
and $W[J]=\ln Z[J]$. (Here $S_E$ is related to the Minkowski action
$S_M=iS_E$
and one could require $S^\dagger_M=S_M$.) The vielbein (\ref{12})
obtains as
\begin{equation}\label{23}
E^m_\mu(x)=\frac{\delta W}{\delta J^\mu_m(x)}
\end{equation}
and obeys the gravitational field equations which follow from the
variation of the effective action $\Gamma$, i.e.
\begin{eqnarray}
\Gamma[E^m_\mu]&=&-W[J]+\int dx\,J_m^\mu(x)E^m_\mu(x),\label{24}\\
\frac{\delta \Gamma}{\delta E^m_\mu(x)}&=&0.\label{25}
\end{eqnarray}
We emphasize that eq. (\ref{25}) constitutes the quantum gravitational
equation. All quantum fluctuations of spinor gravity are already
incorporated into the computation of the effective action $\Gamma$ -
which is, of course,  the difficult and challenging task. The above
construction can be extended to other bosonic and fermionic sources
and fields such that $\Gamma$ contains all information about the
correlation functions of interest.

If $S$ and the regularization of the functional measure $\int {\cal
  D}\psi$ in eq. (\ref{22}) preserve the symmetries of diffeomorphisms
and local Lorentz transformations the effective action $\Gamma$ is
invariant. Expanding in derivatives of the vielbein, $\Gamma$  is
dictated by the symmetries to take the familiar form

\begin{equation}\label{26}
\Gamma=\int d^{16}x\det(E^m_\mu)(c_1+c_2 R+\ldots),
\end{equation}
with $R$ the curvature scalar formed from the metric (\ref{15}) and
$(\det(E_\mu^m))^2=-\det(g_{\mu\nu})$. Dots denote higher invariant
powers of the curvature tensor and its covariant derivatives. This is
a familiar 16-dimensional action for the metric with Einstein and
cosmological constant term.

For a realistic model the static ``ground
state'' solution of the gravitational field equation (\ref{25},\ref{26}) should
preserve the Poincar\'{e} transformations acting on $x^0$ and three
``spatial'' coordinates $(x^1,x^2,x^3)$, while the remaining 12 other coordinates
could be associated to an internal space with characteristic length
scale of the order of the Planck mass. The isometries of the internal
geometry would show up as gauge symmetries in the dimensionally
reduced effective four-dimensional theory \cite{DeW}. For example, if
nine coordinates form a subspace $S^9$ and two more a subspace $S^2$
the four-dimensional gauge symmetry would consist of an $SO(10)$ grand
unification with an $SO(3)$ generation group. The chirality index
\cite{Wet3}  counting the number of massless chiral fermions could be
non-vanishing, especially if the geometry is ``wharped'' \cite{Rub}
and internal space ``non-compact'' \cite{Wet4}.

In conclusion, we have proposed a model where the asymmetry between
time and space arises as a ground state property rather than being
assumed a priori. In our treatment, there is no difference between
``Euclidean time'' and ``Minkowski time'' - both are described by a
common real time coordinate $x^0$. The ``physical signature'' arises
as a consequence  of the expectation value of the bilinear $\tilde
E^m_\mu$. Analytic continuation from the physical space-time to
Euclidean space-time is achieved by analytically continuing  $E^0_\mu$
from its real physical value to an imaginary value, $E^0_\mu\to i
E^0_\mu$.

We have argued that our model of spinor gravity is an interesting
candidate for a renormalizable theory of quantum gravity. In contrast
to earlier approaches we have realized {\it local} Lorentz
symmetry for a well defined action involving only spinors, i.e. for
$S$ an element of the Grassmann algebra which must be polynomial in
the spinor fields. Our 16-dimensional model also has the potential to
provide a unified picture for the observed gauge and gravitational
interactions. Further steps towards a regularized functional measure
and a computation of the effective action are needed before the search
for the ground state can be attacked realiably.  Along the lines of
\cite{Wet5} our approach may also shed more light on the emergence of
a unitary time evolution.


\end{document}